\documentclass[useAMS,usegraphicx,usenatbib]{mn2e}

\title[Simultaneous detections of sGRBs / GW]{Simultaneous event detection rates by electromagnetic and gravitational wave detectors in the advanced era of LIGO and Virgo}

\author[Siellez, Bo\"er and Gendre]{
K. Siellez \thanks{karelle.siellez@oca.eu}, 
M. Bo\" er, 
and B. Gendre
 \\
ARTEMIS (CNRS/UNS/OCA) UMR 7250, Boulevard de l'Observatoire, BP 4229 F-06304 Nice Cedex 4 France\\
}

\begin{document}

\date{Accepted 2013 October 6 - Received 2013 September 23 - In original form 2013  June 5}

\pagerange{\pageref{firstpage}--\pageref{lastpage}} \pubyear{2012}

\maketitle

\label{firstpage}

\begin{abstract}

We present several estimates of the rate of simultaneous detection of the merging of a binary system of neutron stars in the electromagnetic  and  the gravitational wave domains, assuming that they produce short gamma-ay bursts (GRBs). We have based our estimations on a carefully selected sample of short GRBs corrected from redshift effects. The results presented in this paper are based on actual observation only. In the electromagnetic spectrum, we considered observations by current ({\it Swift} and {\it Fermi}) and future (\textit{LOFT} and \textit{SVOM}) missions. In the gravitational wave domain, we consider detections by the Advanced Virgo instrument alone and the network of both Advanced LIGO and Advanced Virgo. We discuss on the possible biases present in our sample, and how to fix them. For present missions, assuming a detection in the following years, we find that we should observe simultaneously between 0.11 and 0.63 gravitational wave events per year with {\it Swift} and {\it Fermi}, respectively. For future projects (\textit{LOFT} and \textit{SVOM}), we can expect less than one common detection per year. We check the consistency of our results with several previously published rate of detection of gravitational waves.

\end{abstract}


\section{Introduction}

With the advent of Virgo and LIGO in their advanced form, and the probable discovery of gravitational sources, astronomy is facing a major shift in its history. Mankind is indeed entering the era of non-photonic detectors, which can detect and observe gravitational waves (hereafter GWs) and possibly neutrinos \citep[e.g.][]{ace09, har10,  acc12, aar13}. This will open new windows on physical events that are, at the moment, totally closed \citep[for a review, see][]{aba12}. It is however probable, and indeed it is already the case \citep[][]{aas13}, that the first detections will be of faint events, and will anyway need to be observed in the electromagnetic domain, both to enhance the confidence on the GW and/or neutrino detection and to optimize the scientific return of the detection itself. This kind of simultaneous observation is however difficult. Theoretical studies \citep{kan12,nis13} have shown all the difficulty of exploring large error boxes with current electromagnetic instrumentation. Any solution to reduce the error box would strongly help. 

Among potential sources of GWs that could lead to an observable phenomenon during the maximal emission of GWs, the merging of two compact objects is one of the most promising and best modelled \citep{abb08,aba10b}. These events are believed to be associated with a class of gamma-ray bursts (GRBs), namely the short GRBs \citep[hereafter sGRBs;][]{eic89}. Even if other progenitors have been proposed for sGRBs \citep[such as newborn magnetars;][]{ref_magnetar}, the merging of two neutron stars (NSs) in a binary system should produce an extremely intense electromagnetic signal. At the same time, this object has been studied theoretically, and observations have proven that they emit their binding energy \citep{tho87}, theoretically as GWs. The fact that sGRBs are detectable up to very large distance \citep[the largest claimed distance of a sGRB is z = 4.6;][]{pos06} should allow easily a combined detection . Ironically, it is the limited distance at which the GW detectors can perform a detection (the range is 150 Mpc for Advanced Virgo and 200 Mpc for Advanced LIGO) that dramatically reduces the number of such events \citep[see][]{aas13}. sGRBs are rare in the Universe, and the sampled volume is so small that the final detection rate is low \citep{cow12,che13}. Despite these limitations, a few strict estimations of the detection rate of an event simultaneously in both windows, based on actual observation, have been done. Most of the results obtained so far were derived from theoretical modelling and population synthesis hypotheses \citep[e.g.][]{gue06}. \citet{pet13} have composed an estimation rate based on the $\textit{Swift}$ observations. However, as we show in Section \ref{Compar}, their selection of data might bias somewhat their result. The aim of this paper is to strengthen this result, using the most recent observations to estimate that rate. 

The major issue of our work is that there is no {\it possible} solution, to date, to claim that a given GRB is caused by the merging of a binary NS system (BNS). One can only assume, and this will be our main hypothesis in this work, that most sGRBs are caused by this phenomenon. Mergers can occur also from neutron star--black hole (NS--BH) or black hole--black hole (BH--BH) binary systems.  These last two types of events are probably detectable at larger distances by AdV/aLIGO \citep[see][]{pac91, sto13}. Yet the signal they produce in the detectors is more difficult to compute over the full parameter space, and their rate is poorly constrained. In addition, it is believed that the electromagnetic signal they produce is weaker than that of BNS systems, though this issue is still debated \citep[see e.g.][]{ros05, davis05}, and it is unclear whether they produce long or short events. Therefore, in this work, we will consider only the issue of the BNS/sGRB connection. 

The second problem we face is that the definition of sGRBs is entirely empirical \citep{kou93} and has no physical ground: sGRBs last less than 2s {\it in the observer frame} and have harder spectra than lGRBs! This definition has an obvious limitation: a burst that would be classified as short at a given redshift would be classified as long at a larger redshift, because of time dilation and cosmological effects \citep[see][]{koc13}. Conversely, this does not exclude that at least some lGRBs originate from the merger of NSs, or BH--NS systems \citep[see e.g.][]{mhpmvp2009}. In this work, we only address the case of sGRBs and we deal with these limitations using another discriminative method to separate sGRBs and lGRBs.

This paper is organized as follows: we present our selection method and our final sample in Section \ref{Sec_method}; we then use it to derive the local rate of sGRBs in Section \ref{sec_result}, and we deduce the rate of simultaneous detections of electromagnetic/GWs events from NS--NS binaries; in Section \ref{sec_discu}, we discuss our results and their consequences in terms of detectability; we finally conclude in Section \ref{sec_ccl}. In the remaining part of this paper, all errors are quoted at $1\sigma$ when not specifically indicated. We use a standard flat cold dark matter model for the Universe, with $\Omega_{m} = 0.23$ and $H_0 = 71$ km s$^{-1}$ Mpc$^{-1}$. sGRB and lGRB stand for short and long GRBs, respectively, EM for electromagnetic waves, GW for gravitational waves, AdV and aLIGO for the Advanced Virgo and Advanced LIGO interferometers, respectively.

\section{Data selection and methods}
\label{Sec_method}

The selection of an unbiased sample is of paramount importance for the estimation of the rate. The main problem is the scarcity of sGRBs, as  only a handful of GRBs discovered from the \textit{Swift} spacecraft have a redshift estimate. 
Meanwhile, the potential sample of bursts (detected by the \textit{Swift} satellite from 2004 December to 2012 June 12) is of 679 bursts detected, among which 191 have a known redshift. We thus decided to reconstruct a more accurate sample using three different filters.

\subsection{Rest-Frame Duration}

As already stated, the use of the canonical definition of a short burst would lead to link this kind of burst with the redshift, a short burst being confused with a long one in case of high redshift (see Equation \ref{eq1}). As we are interested in the nature of the progenitor, this correlation with redshift has to be removed. We thus decided to use the rest-frame duration as a first criteria, 
\begin{equation}
\label{eq1}
\tau_{90} = \frac{T_{90}}{1 + z},
\end{equation}
where $\tau_{90}$ is the 90\% burst duration in the rest frame. We removed from the raw sample all bursts with $\tau_{90} > 2$ s.

\subsection{Spectral selection}

sGRBs are harder than long ones \citep{kou93}. It is thus tempting to select only the hardest bursts, removing all soft events. Empirically, the spectral model that reproduces best the GRB spectrum is the Band model \citep{ban93}. This model consists of a broken power law smoothly joined at a typical energy, $E_0$. From a practical point of view, selecting hard bursts means to set a limit on $E_0$. However, this would again lead to a link with the redshift of the burst, as the observed $E_0$ value depends on the (1 + $z$) factor. Things are even more complicated by the fact that the \textit{Swift}-BAT instrument has a narrow band, preventing a direct filtering from the spectral parameters. Often the BAT instrument detects only one segment of this model. The power-law segment photon indices are usually named $\alpha$ for the soft segment and $\beta$ for the hard one. Typically, the value of $\alpha$ is of the order of 1.2 and $\beta$ of the order of 2.3 \citep{bar03}. We have assumed that for a hard burst, the BAT would have detected only the soft segment $\alpha$ (i.e. the peak energy is above the BAT high-energy limit). This translate to consider a burst to be hard only if the measured spectral index is lower than 2. We rejected all other events.

\subsection{Presence of a plateau phase}

The last parameter of selection is the plateau phase. This phase has been discovered by {\em Swift} \citep{plateau_decouverte} and could be due to energy injection \citep{plateau_energy}. It could also represent a soft tail of a disguised lGRB. Lastly, magnetar progenitors are known to produce a plateau phase \citep[e.g.][]{met11}. As we are interested in the merging of an NS binary system (where few energy should be available once the merging is done), we prefer to remove all bursts with a plateau phase, assuming they are related to other kinds of progenitors \citep[see however][]{gao13}. This is a conservative criteriON and only less than the half of the candidates that passed the two previous filters survived to this one. Because of that, for some rare bursts where the light curve does not allow to determine if a plateau phase is present or not, we relaxed this criteria and validated these events.

\begin{table}
\begin{minipage}{8.5cm}
\caption{The sample of sGRBs used in this work. See text for details. \label{tab_sGRB}}
\small
\begin{tabular}{lccccc}
\hline 
GRB & Redshift &  Spectral & \multicolumn{2}{c}{Duration (s)} & Plateau \footnote{A cross indicates low-quality data preventing a discrimination on this criteria. See text for details} \\
    &          &  index    & Observed & Rest frame           & \\
\hline
101219A  & 0.718  	 &$0.63  \pm 0.09 $ & 0.60  & 0.35 & NO \\
100816A	 & 0.803	 &$0.73  \pm 0.24 $ & 2.90	& 1.61 & NO \\
100724A	 & 1.288	 &$1.92	 \pm 0.21 $	& 1.40	& 0.61 & NO \\
100206A  & 0.41  	 &$0.66  \pm 0.17 $ & 0.12  & 0.09 & $\times$ \\
100117A	 & 0.920	 &$0.88	 \pm 0.22 $	& 0.30	& 0.16 & NO \\
090809	 & 2.737	 &$1.34	 \pm 0.24 $ & 5.40	& 1.45 & NO \\
090510	 & 0.903	 &$0.98	 \pm 0.20 $ & 0.30	& 0.16 & NO \\
080905A  & 0.122	 &$0.85	 \pm 0.24 $ & 1.00	& 0.47 & NO \\
071020	 & 2.142	 &$1.11	 \pm 0.05 $ & 4.20	& 1.34 & NO \\
070429B	 & 0.904	 &$1.72	 \pm 0.23 $ & 0.47	& 0.25 & NO \\
061217	 & 0.827	 &$0.86	 \pm 0.30 $ & 0.21	& 0.11 & $\times$ \\
060801   & 1.131	 &$0.47	 \pm 0.24 $ & 0.49	& 0.23 & NO \\
060502B	 & 0.287	 &$0.98	 \pm 0.19 $ & 0.13	& 0.10 & $\times $ \\
051221A	 & 0.547	 &$1.39	 \pm 0.06 $ & 1.40	& 0.90 & NO \\
050922C	 & 2.198	 &$1.37	 \pm 0.06 $ & 4.50	& 1.41 & NO \\
050813	 & 1.800	 &$1.28	 \pm 0.37 $ & 0.45	& 0.16 & $\times$ \\
050509B	 & 0.225	 &$1.57	 \pm 0.38 $ & 0.07	& 0.06 & $\times$ \\
\hline	 

\end{tabular}
\end{minipage}
\end{table}

The final sample consists oF 17 events, listed in Table \ref{tab_sGRB} together with their properties.

\section{Detection rate}
\label{sec_result}

Our sample consists of 17 events. Among them, four were not classified as sGRBs using the standard criteria of the observed duration $T_{90}$. At the same time, five canonical sGRBs were removed from it. We note that  only eight bursts are in common with the sample of \citet{pet13}; this is due to their different selection criteria (see below). Fig. \ref{f1} presents the redshift distribution. As it can be seen, and as expected, we inserted high-redshift sGRBs though no event with a redshift larger than 2.75 is present. 

\begin{figure}
  \includegraphics[width=9cm]{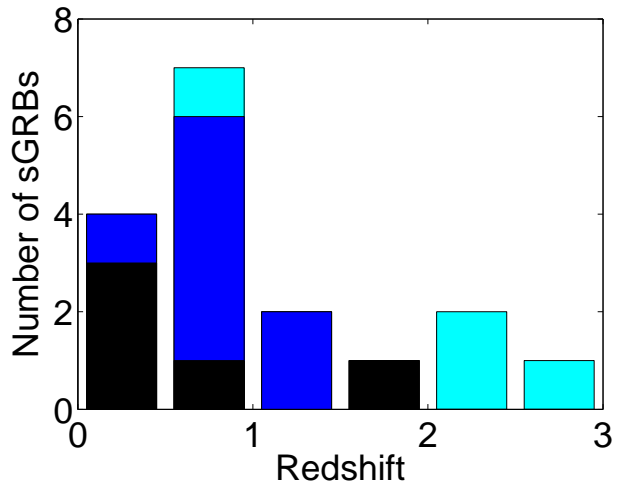}
  \caption{Redshift distribution of the rate of sGRBs by year. We indicate in dark blue the 'classical' short bursts, and in cyan the four events we added. The black ones are canonical sGRBs with no conclusions on the presence of the plateau phase. \label{f1}}
\end{figure}

\begin{figure}
\includegraphics[width=9cm]{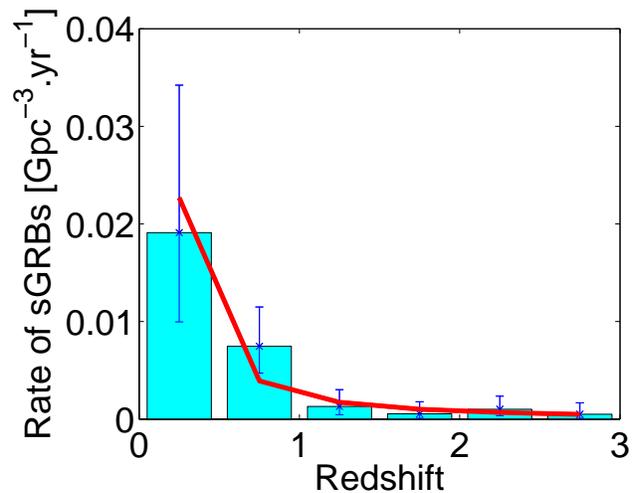}
\caption{The redshift distribution of the rate of sGRBs by year and by comobile volume is in cyan. In dark blue are the error bars using the Poisson statistic. This distribution has been fitted by a power law in red, the numerical form of the event density function is quoted in the text. \label{fig3}}
\end{figure}

From this distribution, we estimated the event density as a function of the redshift, assuming a power-law model with constant rate in the local Universe (i.e. within $z \le$ 0.05), to avoid inconsistency (see below): 
\begin{equation}
\label{power}
Y = a~ z^{b},
\end{equation} 
where $Y$ is the rate of sGRBs by year and by comobile volume.

We obtained a best-fitting power-law index $b$ of $-1.6 \pm 0.5 $, and $ a = 0.0025$ Gpc$^{-3}$ yr$^{-1}$ (Fig. \ref{fig3}).  The correct formula for the detection rate in the Universe is given by  

\begin{equation}
R = \int \int \int_0^{z_{max}} D(z,\theta,\varphi) dz d\theta d\varphi~ \rm{y}^{-1}.
\label{Eq1}
\end{equation}

In equation (\ref{Eq1}), $D$ is the rate of sGRBs per unit volume in the local Universe. It depends on the distance and the direction of the sky considered, because the local Universe is not homogeneous, with the presence of galaxies and voids. Since the catalogues of nearby galaxies located inside the AdV/aLIGO range is far from being complete, we cannot compute equation (\ref{Eq1}) directly.  We add the supplementary hypothesis that the Universe (within that range) is isotropic and homogeneous. This means that $D$ is independent of $\theta$ and $\varphi$ and that $D$ does not vary with the distance. This last statement, is not strictly true: $D$ varies with the distance. Nonetheless, taking its value to be 200 Mpc and assuming it constant is a good proxy for the real integration of its value over the ranges of the instruments. We discuss later the impact of a variation of $D$, i.e. adding close-by or distant GRBs, on our results. This value of $D$ can be obtained by the value of $Y$ given in equation (\ref{power}) for $z=0.05$ corresponding to 200 Mpc and correcting by the fact that $\textit{Swift}$ has a field of view of 1.4 sr instead of 4$\pi$. We obtained:

\begin{equation}
\label{rate}
D = 2.7 \pm 0.9 ~\rm{Gpc}^{-3}~ \rm{yr}^{-1}.
\end{equation}

Applying equations (\ref{Eq1}) and (\ref{rate}), we obtain for Virgo, in its advanced configuration (range of 150 Mpc), $  R = 0.036 \pm 0.012$  yr$^{-1}$. The combination of AdV/aLIGO, which increases the range up to 300 Mpc (Regimbau, private communication), leads to $ R = 0.3 \pm 0.1 $ yr$^{-1}$. These numbers are low, and one may wonder if they are accurate. We discuss this point in the next section, but already note that they are based on a sample detected by {\it Swift}, which is not well suited for detecting sGRBs.

\section{Discussion}
\label{sec_discu}

\subsection{Statistic validation}
\label{Stat_validation}

\begin{table*}
\caption{Summary of our results for the silver sample (see text): we indicate the detection rate density in volume ($D$), the sGRB isotropic event rate ($R$) and the number of simultaneous EM/GW events per year within the field of view of the instrument ($N$) for two different ranges: 150 Mpc (AdV detector) and 300 Mpc (AdV/aLIGO) combined detectors.
See the note added in proof and table 5 for corrected results.\label{table_rate_silver} }
\centering
\small
\begin{tabular}{ccc|ccccc}
\hline 
        &        &              &       & \multicolumn{4}{c}{Horizon} \\
        &        &              &       & \multicolumn{2}{c}{AdV} & \multicolumn{2}{c}{AdV/aLIGO} \\
Mission &  FoV   &  Energy band &    $D$  &         $R$      & $N$      &             $R$   &    $N$        \\
        & (sr)   & (keV)        & (Gpc$^{-3}$ yr$^{-1}$) & (yr$^{-1}$) & (yr$^{-1}$) & (yr$^{-1}$) & (yr$^{-1}$) \\
\hline
{\it Swift} & 1.4 & 15--150   & $9 \pm 3$ & $0.12 \pm 0.04$ & $0.013 \pm 0.005 $& $1.0 \pm 0.4$ & $0.11 \pm 0.05$\\
BATSE   & $\pi$  & 25--1800   & $66 \pm 22$ & $0.87 \pm 0.30$ & $0.22 \pm 0.08$& $ 7.2 \pm 2.4$& $1.8 \pm 0.6$\\
{\it Fermi}-GBM & 9.5 & 8--40000 & $52 \pm 18$   &$0.7 \pm 0.30$ &$0.52 \pm 0.23$ &$5.6 \pm 2.0$ &$4.2 \pm 1.6$\\
{\it LOFT}    & $\pi$  & 2--80      & $36 \pm 12$ &$0.48 \pm 0.14$ & $0.12 \pm 0.04$ & $2.9 \pm 1.0$& $0.7 \pm 0.3$ \\
{\it SVOM}    & 2      & 4--250      & $36 \pm 12$ &$0.48 \pm 0.14$ & $0.08 \pm 0.03$ & $2.9 \pm 1.0$& $0.5 \pm 0.2$  \\
\hline
\end{tabular}
\end{table*}

\begin{table*}
\caption{Same as Table 2 for the gold sample (see text). See the note added in prood and table 6 for corrected results\label{table_rate_gold}}
\centering
\small
\begin{tabular}{ccc|ccccc}
\hline 
        &        &              &       & \multicolumn{4}{c}{Horizon} \\
        &        &              &       & \multicolumn{2}{c}{AdV} & \multicolumn{2}{c}{AdV/aLIGO} \\
Mission &  FoV   &  Energy band &    $D $ &     $    R $     & $N  $    &            $ R $  &   $ N $       \\
        & (sr)   & (keV)        & (Gpc$^{-3}$ yr$^{-1}$) & (yr$^{-1}$) & (yr$^{-1}$) & (yr$^{-1}$) & (yr$^{-1}$) \\
\hline
{\it Swift} & 1.4 & 15--150   & $2.7 \pm 0.9$ & $0.036\pm 0.012 $&$ 0.004 \pm 0.002$ &$ 0.3 \pm 0.1 $ & $0.033 \pm 0.011$ \\
BATSE   & $\pi$  & 25--1800   & $20 \pm 7$ &$0.26 \pm 0.09$ &$0.06 \pm 0.02$ &$2.2 \pm 0.8$ & $0.55 \pm 0.2$ \\
{\it Fermi}-GBM & 9.5 & 8--40000 & $16 \pm 6 $ &$0.21 \pm 0.08$ & $0.16 \pm 0.06$& $1.7 \pm 0.7$ & $1.3 \pm 0.6$\\
\it{LOFT}    & $\pi$  & 2--80      & $11 \pm 4$ & $0.15 \pm 0.06$&$0.04 \pm 0.02$ &$1.2 \pm 0.5$&$0.3 \pm 0.12$ \\
\it{SVOM}    & 2      & 4--250      & $11 \pm 4$ & $0.15 \pm 0.06$&$0.024 \pm 0.003$ &$1.2 \pm 0.5$&$ 0.19 \pm 0.08$  \\
\hline
\end{tabular}
\end{table*}

The range of Advanced Virgo is 0.035 (when expressed in redshift units) and about 0.07 for the network AdV/aLIGO. To date, no sGRB has been detected so close to the Earth. In other words, we have extrapolated the detection rate in a region without data: in the following, we assess the impact of this point. 

\begin{figure}
\includegraphics[width=9cm]{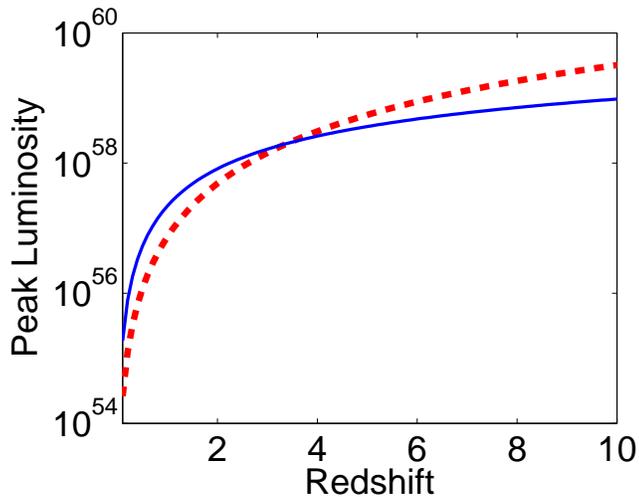}
\caption{Detectability of sGRBs as a function of the redshift. The red dashed line represents the \textit{Swift} detection sensibility for an sGRB with all properties (duration, peak flux, Band parameters) set to the median of the observed values. The blue solid line represents the peak flux of this template GRB. \label{fig2}}
\end{figure}

In Fig. \ref{fig2}, we compare the peak flux of a template sGRB with the detection threshold of \textit{Swift} for this event. As one can clearly see, at low redshift this kind of burst can be detected, while at high redshift there are selection effects at play. 

As the fit using a power law might reflect also an underestimation of the GRB rate at high redshift because of an instrumental bias, we have tested this possibility and the influence of a larger number of distant events: let  us add a burst at large redshift in our sample and recompute all rates. We find that they remain constant within errors. This can easily be explained: at high redshift, the sampled volume is so large that the addition of a few bursts will not be significant. Only a large population of NS--NS systems could modify it; however, the merging of a binary system of NSs is a process that takes a long time to occur \citep{ref_evol_BNS}, and it is possible that this population of binaries is still partly evolving (i.e. not merging) at high redshift. In any case, the impact of the rejection of a burst at high redshift on the local density is negligible.

The situation is different at low redshift. We have again inserted a burst in our sample, this time at redshifts z = 0.04, 0.08 and 0.11 \citep[the latter two values being the proposed redshifts for GRB 061201 not present in our sample; see][]{ber07}. We find that for all cases, the addition of this nearby burst multiply by a factor about 2 the rates we have obtained so far (the largest rate when considering the 300 Mpc range with this modified sample is $R = 0.4 \pm 0.1$ sGRB yr$^{-1}$). Thus, even if an uncertainty is larger in the final rate, our estimates are still valid. 

In the following, we will maintain our initial sample, but we will discuss the implications taking into account this uncertainty.

\subsection{Removing the redshift measurement bias}

Our sample is based only on sGRBs with a measured redshift. Indeed, this could be considered as a 'gold sample' of sGRBs that would be detected in the gamma-ray, X-ray, optical bands (i.e. in the electromagnetic spectrum) as well as with GWs.

Using the canonical definition of an sGRBs : $T_{90} < 2$ s, we find 57 sGRB and among them 18 have a known redshift. We conclude that only 31.6\% of sGRBs detected by {\em Swift} have a redshift measurement. We thus  define a "silver sample" of sGRBs that will be detected simultaneously in EM and GW without an associated redshift measurement (e.g. because the afterglows fade quickly, of a dark GRBs, faint sources, no confirmed host, etc.). We note that this situation can change dramatically with the recent discovery of a probable kilonova associated with an sGRB \citep{ ber13, tan13}.

Our method to select sGRBs is based on the redshift measurement (to get the $\tau_{90}$ value).  We assume that the ratio of sGRB without redshift to the ones with redshift measurement is the same as for canonical sGRBs (31.6 per cent). We also assume that these sGRBs have the same redshift distribution as our 'gold sample'. This second hypothesis is fair, as most of the sGRBs are nearby, where selections effects play no role when estimating the redshift \citep{cow13}. Using these numbers, we find that the rate for the silver sample is $ R=0.12 \pm 0.04$ yr$^{-1}$ for a 150 Mpc range (AdV) and $R=1.0 \pm 0.4$ yr$^{-1}$ for a range of 300 Mpc (aLIGO/AdV combined).

\subsection{Removing the instrumental bias}

{\it Swift} is not the best-suited instrument to detect sGRBs \citep[see e.g.][]{zha07}. In fact, the BATSE 4B catalog \citep{pac99} contains a larger proportion of sGRBs. There is thus another bias to correct, linked to the sensitivity of the instrument. We have assumed that the discrepancy in sensitivity does not modify the distribution in redshift nor the ratio of sGRBs selected with our method to canonical sGRBs. This last statement means that this ratio, equal to 18/57 (see previous section), is constant for all missions. We are then able to reconstruct the gold and silver samples for past and present missions. The detection rate density in volume $D_{inst}$ for each instrument is calculated following this formula\footnote{See Section 'NOTE ADDED IN PROOF' for more accurate results.} :

\begin{equation}
D_{inst} = D_{\textit{Swift}} \times \frac{(number~of~sGRBs  / yr)_{inst}}{(number~of~sGRBs/yr)_{\it{Swift}}}
\end{equation}

$D_{\textit{Swift}}$ is $D$ expressed for \textit{Swift} in Tables \ref{table_rate_silver}\footnotemark[1] and \ref{table_rate_gold}\footnotemark[1] and the number of sGRBs per year for \textit{Swift} is 7.6. The number of sGRB per year for each instrument is obtained either with the estimation of expected trigger number given by the instrument collaboration [see \citealt{fer12} for the LOFT mission, and Atteia, priv. com. for SVOM] or from the published catalogues. These values are 44 sGRB per year for the {\em Fermi}-GBM detector, using their online catalogue (from 2008 July to 2010 July) and 55.6 for BATSE using the 4B catalog. The results are given in Tables \ref{table_rate_silver} and \ref{table_rate_gold} for the silver and gold samples, respectively.

\subsection{Best observation strategy}

\begin{table*}
\caption{Predictions of GW detection rates by comobile volume from this work and comparison with other authors: the first column gives the paper reference, the second the method used by the others and the last the estimated rate \label{table_angle}}
\centering
\small
\begin{tabular}{ccc}
\hline 
Work               & Method   &  Estimated GW detection rate \\
                   &          & (Gpc$^{-3}$ yr$^{-1}$)\\
\hline
This work						  & Observational constraints & 92--1154 \\
\citet{cow12}                             & Observational constraints & 8--1800 \\
\citet{pet13}							  & Observational constraints & 500--1500 \\
\citet{gue06}                             & Theoretical modelling      & 8--30 \\
\citet{aba10}                             & Theoretical modelling      & 2.6--2600 \\
\hline
\end{tabular}
\end{table*}

The estimated rates listed in Tables \ref{table_rate_silver}\footnotemark[1] and \ref{table_rate_gold}\footnotemark[1] are valid for the whole sky. Because the different ranges of GW detectors correspond to a volume of 0.013--0.108 Gpc$^3$, the final numbers are low. For instance, considering {\it Swift}, the final number of common EM/GW events that can be expected each year is 0.11. In the best possible scenario, i.e. with {\it Fermi}-GBM, we obtain\footnotemark[1] $N = 4.2 \pm 1.6$ sGRB yr$^{-1}$ for a 300 Mpc range corresponding to the combination of AdV/aLIGO (Regimbau, private communication). This means that only 2--3 events\footnotemark[1] per year should lead to an observation simultaneously at high energy and in GW. The GBM uncertainties on the GRB positions are large (of the order of 100deg$^2$). It is thus a key point to be prepared to observe a large portion of the sky with enough sensitivity. The use of optical telescopes with a wide field of view such as TAROT \citep{ref_tarot,klo09} will be critical. Radio instruments such as  SKA and LOFAR may also scan a whole error box at a glance, and indeed working groups are already preparing the follow-up of EM/GW transients \citep[see e.g.][]{ref_VAST}.

\subsection{Comparison with other results}
\label{Compar}

Previous works have in general not derived the rate of dual observations, rather the rate of detection of GWs. Two main methods have been followed: population synthesis models evolved up to the merging of NS binaries \citep[e.g.][]{gue06}, and observed sample of short bursts corrected for selection effects \citep[e.g.][]{cow12}. Both methods have their advantages and disadvantages, but give consistent results of a few tens of GW triggers per year. We now check if our findings are consistent with these results.

Based on observations, our method is already corrected for several biases and gives the actual common EM/GW detection rate. In order to convert it to the GW detection rate, we have to apply a correction for the jet opening half-angle $\theta_{j}$ given by \citep{rho99} 

\begin{equation}
B(\theta_{j}) = [1 - cos(\theta_{j})]^{-1}.
\label{angle}
\end{equation}

The main uncertainty is the value of $\theta_j$ one should use. The only measurement of it in our sample is $\theta_j = 7^\circ$ for GRB 051221A \citep{sod06}. Using this value for all bursts, we obtain $D_{GW} = 1154 \pm 389$ Gpc$^{-3}$ yr$^{-1}$. However, \citet{cow11} indicate that the beaming angle derived for GRB 051221A is the lowest measured for any sGRBs and may not be representative. Hence, they use a larger value of $\theta_{j} = 14^\circ$. When we use this last number, we obtain $ D_{GW} = 290 \pm 98 $ Gpc$^{-3}$ yr$^{-1}$. The largest measured value of $\theta_{j}$ is $\sim~25^\circ$ \citep{gru06}. Using this limit, we found  $ D_{GW} = 92 \pm 31 $ Gpc$^{-3}$ yr$^{-1}$. Hence, our estimation of $D_{GW}$ is between 92 and 1154.

We reported in table \ref{table_angle} all published values of the GW detection rate estimation. As one can see, we are in agreement with all but \citet{gue06}. These authors, using the population synthesis method, find a rate between 8 and 30 events Gpc$^{-3}$ yr$^{-1}$. These values are excluded at more than 2$\sigma$; we are in disagreement with this estimate.

As stated in Section \ref{sec_result}, we have a somewhat different sample than that of \citet{pet13}. Nevertheless, the final results are in agreement. This can straightforwardly be explained by our studies reported in Section \ref{Stat_validation}. The addition of a few bursts does not change significantly the results. As our sample have roughly the same size, the results must be similar. We note that our lower limit is lower. When translated this into simultaneous detection rate, they found a number of events per year between 0.2 and 1, while we found a simultaneous detection rate between 0.06 and 0.16 events per year, 10 times lower. This discrepancy can be explained by the use of a larger, calibrated sample in our case, where 'fake' sGRBs were removed  \citep[e.g.][]{zha09, bro13}. In addition, \citet{pet13} have restricted their sample in date and redshift: they  neglected sGRBs located in the redshift desert, while this region is very important for a global census \citep{cow13}.  \citet{pet13} have also reduced there sample by 15 \% to take into account possible magnetars: however, the magnetar should produce a plateau phase that can be used to discriminate lGRBs from sGRBs \citep{dal11}. Their sample is thus restricted to 14 bursts, only. In addition, we choose a more conservative horizon for the detectors.

 We consequently conclude that our estimates are fair and in good agreement with previous papers. Again, we emphasize that our work is based on observational constraints rather than on theoretical computations.

\section{CONCLUSION}
\label{sec_ccl}

In this paper, we have presented an estimate of the rate of simultaneous detection of sGRBs and GW events, assuming that they originate from the same event, namely the coalescence of an NS--NS binary system. We used the {\it Swift} catalogue to derive a set of 17 sGRBs corrected from instrumental/local effects. This sample has been used to derive the rate density of events expected from present and future GRB missions ({\it Swift}, {\it Fermi}, {\it LOFT} and {\it SVOM}) within the range of Advanced Virgo and of the combination of Advanced Virgo/Advanced LIGO.

While the rate of common EM/GW detection for which we can expect that a redshift will be measured (assuming that {\it Swift} will still be in operation) will be low (about 0.03yr$^{-1}$), we expect a fair number of events simultaneously detected by {\it Fermi} and AdV/aLIGO, i.e. close to 1.5\footnote[1]{See Section 'NOTE ADDED IN PROOF' for more accurate results.} common detections per year.
We defined two samples, one gold sample that should be observed at all wavelengths (i.e. with a redshift estimate), and a second, silver, sample of events detected only in gamma-ray and by GWs.  

These numbers, even if not high, are large enough to allow a confirmation of the detection of GWs during the first years of operation of the instruments, and common study of the sources with both EM and GW radiations. Planned missions (\textit{LOFT} and \textit{SVOM}) will not increase this rate, and in fact {\it Fermi} is more suited for this task due to its larger field of view and higher sensitivity and energy range.

The construction of the advanced versions of Virgo and LIGO has already started and the first scientific runs have been scheduled for 2015 (LIGO) and 2017 (Virgo). The Japanese detector, KAGRA, is also on its way, and the INDIGO (India) project has been approved. It is therefore of paramount importance to optimize the scientific return of these large experiments. The EM follow-up is a way to both confirm a detection (especially if the confidence based on GW only is low), and to maximize the science that can be done and the understanding of the sources (NSs) as well as the dynamics of the coalescing binary system and its by-product (the sGRB). Preparing a comprehensive set of EM instruments at all wavelengths, encompassing radio, IR, optical, X-ray, gamma-rays is an important objective that should be addressed before Virgo and LIGO start their operational life, i.e. now.

\section*{Acknowledgements}

We thank the anonymous referee for helpful comments and suggestions We acknowledge the use of public data from the \textit{Swift} data archive. We thank Gianluca Gemme for useful comments and suggestions.

\section*{Note Added in Proof}

We realized that there is an error in equation 5, which should be read as rate density in volume per steradian: therefore the units are sGRBs/yr/sr. This does not change the estimates of common EM/GW events for \textit{Swift}. 

For the other instruments the rates N for the silver sample (Table 2) become $0.8 \pm 0.3$, $0.63 \pm 0.21$, $0.14 \pm 0.05$ and $0.14 \pm 0.05$ for BATSE, \textit{Fermi}-GBM, \textit{LOFT} and \textit{SVOM} respectively, see Table \ref{table_rate_silver_modif}. For the gold sample (Table 3) the rates N become $0.24 \pm 0.08$, $0.19 \pm 0.07$, $0.04 \pm 0.01$ and $0.04 \pm 0.01$ for BATSE, \textit{Fermi}-GBM, \textit{LOFT} and \textit{SVOM} respectively, see Table \ref{table_rate_gold_modif}.

This means that we expect about 1--2 common detections between \textit{Fermi}-GBM and AdV/ALIGO combined over their nominal operational life (3 years). 

We thank Neil Gehrels for having pointed out this problem in our paper.

\label{lastpage}

\newpage

\begin{table*}
\caption{Summary of our results for the silver sample with more accurate results (see Section 'NOTE ADDED IN PROOF'): we indicate the detection rate density in volume ($D$), the sGRB isotropic event rate ($R$) and the number of simultaneous EM/GW events per year within the field of view of the instrument ($N$) for two different ranges: 150 Mpc (AdV detector) and 300 Mpc (AdV/aLIGO combined detectors). \label{table_rate_silver_modif}}
\centering
\small
\begin{tabular}{ccc|ccccc}
\hline 
        &        &              &       & \multicolumn{4}{c}{Horizon} \\
        &        &              &       & \multicolumn{2}{c}{AdV} & \multicolumn{2}{c}{AdV/aLIGO} \\
Mission &  FoV   &  Energy band &   $ D $ &  $       R     $ & $N$      &       $      R $  & $   N $       \\
        & (sr)   & (keV)        & (Gpc$^{-3}$ yr$^{-1}$) & (yr$^{-1}$) &( yr$^{-1}$) & (yr$^{-1}$) & (yr$^{-1}$) \\
\hline
{\it Swift} & 1.4 & 15--150   & $9 \pm 3$ & $0.12 \pm 0.04$ & $0.013 \pm 0.004 $& $1.0 \pm 0.3$ & $0.11 \pm 0.04$\\
BATSE   & $\pi$  & 25--1800   & $30\pm 10$ & $0.4 \pm 0.1$ & $0.10\pm 0.03$& $ 3.2\pm 1.1$& $0.80\pm 0.30$ \\
{\it Fermi}-GBM & 9.5 & 8--40000 & $7.7\pm 2.6$   &$0.1\pm 0.03$ &$0.08\pm 0.03$ &$0.8\pm 0.3$ &$0.63\pm 0.21$\\
{\it LOFT}    & $\pi$  & 2--80      & $5.3\pm 1.8$ &$0.07\pm 0.02$ & $0.02\pm 0.01$ & $0.6\pm 0.2$& $0.14\pm 0.05$ \\
{\it SVOM}    & 2      & 4 --250      & $8.3\pm 2.8$ &$0.11\pm 0.04$ & $0.02\pm 0.01$ & $0.9\pm 0.3$& $0.14\pm 0.05$  \\
\hline
\end{tabular}
\end{table*}

\begin{table*}
\caption{Same as Table \ref{table_rate_silver_modif} for the gold sample with more accurate results (see Section 'NOTE ADDED IN PROOF'). \label{table_rate_gold_modif}}
\centering
\small
\begin{tabular}{ccc|ccccc}
\hline 
        &        &              &       & \multicolumn{4}{c}{Horizon} \\
        &        &              &       & \multicolumn{2}{c}{AdV} & \multicolumn{2}{c}{AdV/aLIGO} \\
Mission &  FoV   &  Energy band &    $D$  &     $    R$      & $N  $    &   $          R  $ &$    N $       \\
        & (sr)   & (keV)        & (Gpc$^{-3}$ yr$^{-1}$) & (yr$^{-1}$) & (yr$^{-1}$) & (yr$^{-1}$) & (yr$^{-1}$) \\
\hline
{\it Swift} & 1.4 & 15--150   & $2.7 \pm 0.9$ & $0.04\pm 0.01$&$ 0.004 \pm 0.001$ &$ 0.3 \pm 0.1 $ & $0.03 \pm 0.01$ \\
BATSE   & $\pi$  & 25--1800   & $8.8\pm 2.9$ &$0.12\pm 0.04$ &$0.030 \pm 0.010$ &$1.0 \pm 0.3$ & $0.24 \pm 0.08$ \\
{\it Fermi}-GBM & 9.5 & 8--40000 & $2.3 \pm 0.8$ &$0.03 \pm 0.01$ & $0.023 \pm 0.008$& $0.2 \pm 0.1$ & $0.19 \pm 0.07$\\
{\it LOFT}    & $\pi$  & 2--80      & $1.6 \pm 0.5$ & $0.02 \pm 0.01$&$0.005 \pm 0.002$ &$0.2 \pm 0.1$&$0.04 \pm 0.01$ \\
{\it SVOM  }  & 2      & 4--250      & $2.5 \pm 0.8$ & $0.03 \pm 0.01$&$0.005 \pm 0.002$ &$0.3 \pm 0.1$&$ 0.04 \pm 0.01$  \\
\hline
\end{tabular}
\end{table*}

\end{document}